\documentclass[prl,floatfix,superscriptaddress,nofootinbib,showpacs,showkeys,preprintnumbers,twocolumn]{revtex4}

\usepackage[dvips]{epsfig}
\usepackage{graphicx}

\begin{document}

\preprint{PREPRINT NPI MSU 2005-22/788}

\title{Further study of narrow baryon resonance decaying into $K^0_s p$ in $pA$-interactions at $70\ GeV/c$ with SVD-2 setup.}

\def\groupsinp{\affiliation{D.V. Skobeltsyn Institute of Nuclear Physics, Lomonosov Moscow State University, 1/2 Leninskie gory, Moscow, 119992 Russia}}
\def\groupihep{\affiliation{Institute for High-Energy Physics, Protvino, Moscow oblast,  142284, Russia}}
\def\groupjinr{\affiliation{Joint Institute for Nuclear Research, Dubna, Moscow oblast, 141980, Russia}}
\def\groupzel{\affiliation{Research Institute of Materials Science and Technology, 103460, Moscow, Zelenograd, Russia}}
\def\groupzell{\affiliation{NPO "NIITAL",  103460, Moscow, Zelenograd, Russia}}

\groupsinp
\groupihep
\groupjinr
\groupzel
\groupzell

\author{A.~Aleev} \groupjinr
\author{E.~Ardashev} \groupihep
\author{V.~Balandin} \groupjinr
\author{S.~Basiladze} \groupsinp
\author{S.~Berezhnev} \groupsinp
\author{G.~Bogdanova} \groupsinp
\author{I.~Boguslavsky} \groupjinr
\author{V.~Bychkov} \groupjinr
\author{N.~Egorov} \groupzel
\author{V.~Ejov} \groupsinp
\author{G.~Ermakov} \groupsinp
\author{P.~Ermolov} \groupsinp
\author{N.~Furmanec} \groupjinr
\author{V.~Golovkin} \groupihep
\author{S.~Golovnia} \groupihep
\author{S.~Golubkov} \groupzel
\author{A.~Gorkov} \groupzell
\author{S.~Gorokhov} \groupihep
\author{I.~Gramenitsky} \groupjinr
\author{N.~Grishin} \groupsinp
\author{Ya.~Grishkevich} \groupsinp
\author{D.~Karmanov} \groupsinp
\author{A.~Kholodenko} \groupihep
\author{A.~Kiriakov} \groupihep
\author{N.~Kouzmine} \groupjinr
\author{V.~Kozlov} \groupsinp
\author{Yu.~Kozlov} \groupzel
\author{E.~Kokoulina} \groupjinr
\author{V.~Kramarenko} \groupsinp
\author{A.~Kubarovsky\footnote[1]{Contact person, e-mail: alex\_k@hep.sinp.msu.ru}} \groupsinp
\author{L.~Kurchaninov} \groupihep
\author{V.~Kuzmin} \groupsinp
\author{E.~Kuznetsov} \groupsinp
\author{G.~Lanshikov} \groupjinr
\author{A.~Larichev} \groupsinp
\author{A.~Leflat} \groupsinp
\author{M.~Levitsky} \groupihep
\author{S.~Lyutov} \groupsinp
\author{M.~Merkin} \groupsinp
\author{A.~Minaenko} \groupihep
\author{G.~Mitrofanov} \groupihep
\author{V.~Murzin} \groupsinp
\author{V.~Nikitin} \groupjinr
\author{P.~Nomokonov} \groupsinp
\author{S.~Orfanitsky} \groupsinp
\author{V.~Parakhin} \groupihep
\author{V.~Petrov} \groupihep
\author{L.~Pilavova} \groupzell
\author{V.~Peshekhonov} \groupjinr
\author{A.~Pleskach} \groupihep
\author{V.~Popov} \groupsinp
\author{V.~Riadovikov} \groupihep
\author{R.~Rudenko} \groupihep
\author{I.~Rufanov} \groupjinr
\author{V.~Senko} \groupihep
\author{N.~Shalanda} \groupihep
\author{A.~Sidorov} \groupzel
\author{M.~Soldatov} \groupihep
\author{L.~Tikhonova} \groupsinp
\author{T.~Topuria} \groupjinr
\author{Yu.~Tsyupa} \groupihep
\author{A.~Uzbyakova} \groupsinp
\author{M.~Vasiliev} \groupihep
\author{A.~Vischnevskaya} \groupsinp
\author{K.~Viriasov} \groupjinr
\author{V.~Volkov} \groupsinp
\author{A.~Vorobiev} \groupihep
\author{A.~Voronin} \groupsinp
\author{V.~Yakimchuk} \groupihep
\author{A.~Yukaev} \groupjinr
\author{L.~Zakamsky} \groupihep
\author{V.~Zapolsky} \groupihep
\author{N.~Zhidkov} \groupjinr
\author{V. ~Zmushko} \groupihep
\author{S.~Zotkin} \groupsinp
\author{D.~Zotkin} \groupsinp
\author{E.~Zverev} \groupsinp

\collaboration{The SVD Collaboration} \noaffiliation

\date{\today. ~To be submitted to Yadernaya Fizika}


\begin{abstract}
The inclusive reaction $p A \rightarrow pK^0_s + X$ was studied at IHEP accelerator with $70\ GeV$ proton beam using SVD-2 detector.
Two different samples of $K^0_s$, statistically independent and belonging to different phase space regions were used in the analyses and a narrow baryon resonance with the mass $M=1523\pm 2(stat.)\pm 3(syst.)\ MeV/c^2$ was observed in both samples of the data. The statistical significance was estimated to be of $8.0~\sigma$ (392 signal over 1990 background events). Using the part of events reconstructed with better accuracy the width of resonance was estimated to be $\Gamma < 14~MeV/c^2$ at 95\% C.L.
\end{abstract}

\maketitle

\section {Introduction.}
In the last two years the observation of a narrow baryon state named
$\Theta^+$ predicted by Diakonov, Petrov and Polyakov\cite{dpp} has been reported by a large number of experiments in the
the $nK^+$ or $K^0_s p$ decay channels\cite{nakano,nakano2, diana,clas1,saphir,itep,clas2,hermes,svd4,zeus,nomad}. However, several experiments, mostly at high energies, did not confirm the existence of $\Theta^+$. The complete list of references to positive and negative results with discussion can be found in \cite{dzierba,hicks,danilov}. The situation is getting more intriguing, as recently CLAS collaboration reported negative results on $\Theta^+$ photoproduction off proton and deutron with high statistics\cite{burkert}. Meanwhile LEPS collaboration reported the new evidence of $\Theta^+$-baryon in the reaction $\gamma d\rightarrow \Theta^+ \Lambda^*(1520)$\cite{nakano3}. Also, STAR collaboration observed the doubly charged exotic baryon in the $pK^+$ decay channel\cite{star}. Therefore, pentaquark existence is still under the question and new experiments are needed to confirm or refute it.
The SVD-2 collaboration reported the observation of narrow baryon resonance in the $K^0_s p$-system with the mass of $M=1526\pm3(stat.)\pm 3(syst.)~MeV/c^2$ and $\Gamma < 24~MeV/c^2$\cite{svd4}.
In that analysis the $K^0_s$-mesons decayed inside vertex detector (decay length $\le 35\ mm$) were used.
In this paper we present the new results for the study of the same reaction:\\
$pN\rightarrow \Theta^+ + X$, \ $\Theta^+ \rightarrow pK^0_s$, \  $K^0_s \rightarrow \pi^+\pi^-$\\
\noindent
We have used two independent data samples, selected by the point of $K^0_s$ decay: inside or outside the vertex detector ( decay length $0 - 35$ or $35 - 600\ mm$, respectively).

\begin{center}
\begin{figure}[ht]
\vspace{70mm}
{\includegraphics{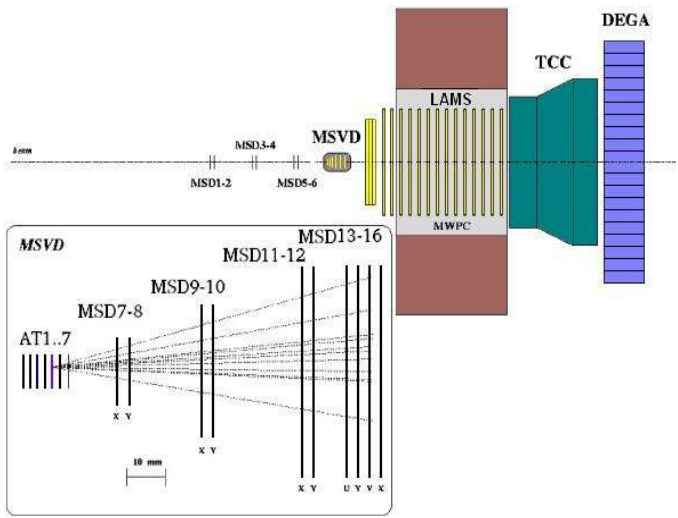}}
\caption{SVD-2 layout.}
\label{setup}
\end{figure}
\end{center}

\section {SVD-2 experimental setup.}

A detailed description of SVD-2 detector and the trigger system can be found elsewhere\cite{svd4,svdtrig}. A brief outline of the main components most relevant to this analysis is given below (Fig.\ref{setup}).

\begin{enumerate}
\item The high-presicion microstrip vertex detector(MSVD) with active(Si) and passive(C,Pb) nuclear targets(AT)
\item Large aperture magnetic spectrometer(LAMS).
\item The multicell threshold Cherenkov counter(TCC).
\item The gamma quanta detector (DEGA).
\end{enumerate}

Data taking was performed in the proton beam of IHEP accelerator with energy $E_p = 70\ GeV$ and intensity $I \approx (5 \div 6) \cdot 10^5$ p/cycle. The total statistics of $5\cdot10^7$ inelastic events was obtained. The sensivity of this experiment for inelastic $pN$-interactions taking in account the triggering efficiency was $1.6~nb^{-1}$.

\section {Analysis I: $K^0_s$ decaying inside the vertex detector (decay length is $0 - 35\ mm$).}

Since SVD-2 reported a pentaquark observation\cite{svd4} (January, 2004) the new improved algorithms of the tracks reconstruction
have been developed \footnote {This part of work was done by one of us (V. Volkov)}. It allowed to increase the number of $K^0_s$ decaying inside vertex detector in 3-4 times and the experimental $K^0_s$ mass resolution for this area was essentially improved. Identification of protons was used for particles in the aperture of TCC, outside it any positively charged track was considered as proton in $K^0_s p$ mass combination.
The first step in the  analysis was to find out events with a well defined secondary vertex at the distance of 0 - 35 mm of the beam direction from the beginning  of the active target (corresponding to the sensitive area of the vertex detector). Tracks from secondary vertex were explored to the magnetic spectrometer and their momenta were reconstructed.

\begin{figure}[ht]
\vspace{80mm}
{\includegraphics{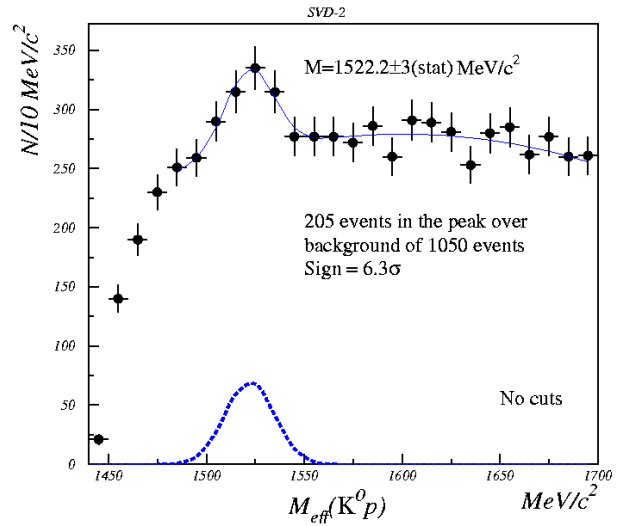}}
\caption{Analysis I: The $(pK^0_s)$ invariant mass spectrum for $K^0_s$ decaying inside the vertex detector without any cuts.}
\label{vfig4}
\end{figure}

This improved analysis allowed to obtain the mass resolutions to be $3.98~MeV/c^2$ and $1.46~MeV/c^2$ for $K^0_s$ and $\Lambda$ masses, respectively. The events with about 3$\times$10$^4$ kaon candidates with the multiplicity $\leq$ 7 were taken for the further processing.
Next step was to reconstruct momenta of the particles from the primary
vertex and to build the invariant mass distributions using previously
selected neutral particles in association with primary vertex tracks. Using (K$^{0}$$\pi^{+}$) and ($\Lambda$$\pi^{+}$) invariant mass spectra it was confirmed that the masses and widths of K$^{*}$(892)$^{+}$-meson and $\Sigma$(1385)$^{+}$-baryon are consistent with PDG tables\cite{pdg}.

\begin{figure} [h]
\vspace{95mm}
{\includegraphics{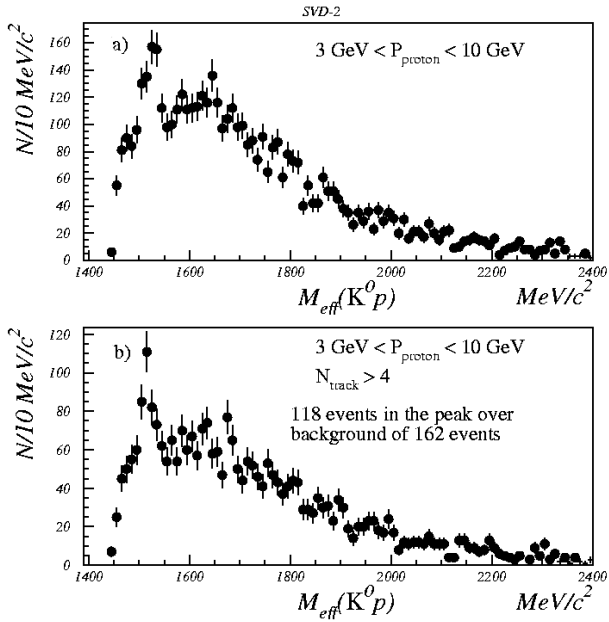}}
\caption{Analysis I: The $(pK^0_s)$ invariant mass spectrum for $K^0_s$ decaying inside the vertex detector with the cuts explained in text.}
\label{vfig3}
\end{figure}

The $K^0_s p$ invariant mass spectrum is shown in Fig.\ref{vfig4}.
Clear excess is observed in the area of interest with the
significance of about 6.2$\sigma$ (205 signal over 1050 background events). Then, the following cuts have been applied:

\begin{itemize}
  \item
3 GeV $<$ P$_{proton}$ $<$  10 GeV , \\where  P$_{proton}$  is the
momentum of the associated primary vertex track. It decreases the background from the kinematically forbidden tracks (the low momentum pions from K$^{0}$ decay are inaccessible with the SVD-2 spectrometer) (Fig.\ref{vfig3}a);
  \item
N$_{tracks}$ $>$ 4,\\
     where  N$_{tracks}$ is the number of tracks associated with the primary vertex (Fig.\ref{vfig3}b).
\end{itemize}

As a result, a stronger peak over the smooth background has been appeared. In the window of $1500 - 1540~MeV/c^2$ there are 118 events over 162 events of background, with the Gaussian $\sigma$ of the peak of about $8~ MeV/c^2$. The statistical significance of the peak can be calculated with different estimators: \\
s/$\sqrt{b}$ = 9.2$\sigma$, s/$\sqrt{s + b}$ = 7.2$\sigma$,
s/$\sqrt{s + 2b}$ = 5.6$\sigma$,

\noindent
where s and b are the numbers of signal and background events respectively.
These values are better than any other published results. The first estimator seems to be more suitable in our case as we estimate the
probability of the statistical fluctuation over the smooth background. The matter of dicussion is an estimate of the background itself in the case when there are different ways to build it.

\begin{figure}[ht]
\vspace{80mm}
{\includegraphics{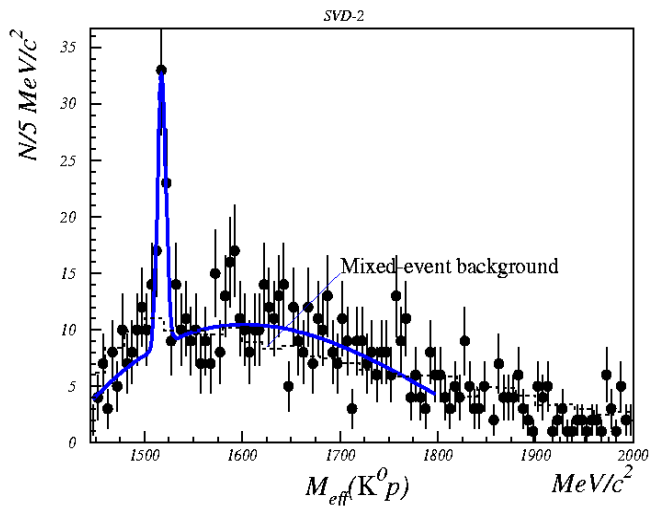}}
\caption{Analysis I: The $(pK^0_s)$ invariant mass spectrum for $K^0_s$ decaying inside the vertex detector with additional quality cuts explained in text.}
\label{vfig5}
\end{figure}

To estimate the natural width of the observed peak it is necessary to select events with the best reconstruction accuracy. Since the experimental resolution of SVD-2 setup (to some degree) depends on geometric parameters of tracks the quality cuts have been applied in order to examine the real intrinsic width of the signal:

\begin{itemize}
  \item
D $<$ 3$\sigma$, \\where  D is the distance of the closest approach
between $V_0$ tracks taken in standard deviations;
  \item
N$_{hits}$ $\geq$ 12, \\where N$_{hits}$ is the number of hits on
the track in the magnetic spectrometer.
\end{itemize}

The result is shown in Fig.\ref{vfig5}. Here the dotted curve
stands for a normalized mixed-event background where kaon and
proton are taken from different events with the same kinematic
requirements. The distribution was fitted by the sum of Gaussian function and second-order polynomial background. We have 50 events in the peak over 25 background events in the mass window of $1510 -
1526~MeV/c^2$ with the statistical significance of about 10$\sigma$.
The narrow Gaussian width of $4.1 \pm 1.0~MeV/c^2$ is consistent with our estimate of the highest possible experimental resolution of the SVD-2 setup. Taking into account the experimental resolution of the SVD-2 detector (calculated using well-known resonances) it was estimated that intrinsic width of the $(pK^0_s)$-resonanse is $\Gamma < 14~MeV/c^2$ at 95\% C.L.

The shape of mixed-event background excludes the
possibility of generating the sharp peak due to the
applied cuts and the detector acceptance. The comparison between the combinations with positive and negative tracks shows no signal in the wrong charge combinations. In addition, possible kinematic reflections (K$^*$(892)$^+$ $\rightarrow$ M(K$^{0}$p) and $\Sigma$(1385)$^+$ $\rightarrow$ M(K$^{0}$p)) and mechanisms with ghost tracks described in\cite{longo} were checked out and the null result was obtained. On
the other hand, it is very difficult to obtain a narrow strong peak generating due to any kind of reflections or
kinematic constraints on the data.

\section {Analysis II: "Distant" $K^0_s$ decaying outside the vertex detector (decay length is $35 - 600\ mm$).}

In Analysis II $K^0_s$-mesons and $\Lambda$-hyperons were reconstructed by the following procedure. First, events with $n_{ch}\le8$ in the primary vertex were reconstructed using the algorithm described in\cite{svd4} and the spectrometer hits belonging to the reconstructed tracks were removed. The remaining spectrometer hits were used for reconstructing track candidates which may originate from secondary vertices in the region before the first spectrometer plane (decay region $35 - 600\ mm$). The initial track candidate had to have at least 3 hits in the Y-planes which can be approximated by a polynomial function and originates from the area of interest. After that additional hits in the U and V planes of the spectrometer were searched for and global track parameters were defined using magnetic field map. Opposite charged tracks then were combined to test whether they may originate from a common secondary vertex. Based on the intrinsic tracking resolution of the spectrometer the minimum distance between two tracks was required to be less than 1 mm in the horizontal and 5 mm in the vertical directions. The efficiency of this method is about 50\% and it's lower than for the
methods of reconstruction using vertex detector information.

\begin{figure}[ht]
\vspace{100mm}
{\includegraphics{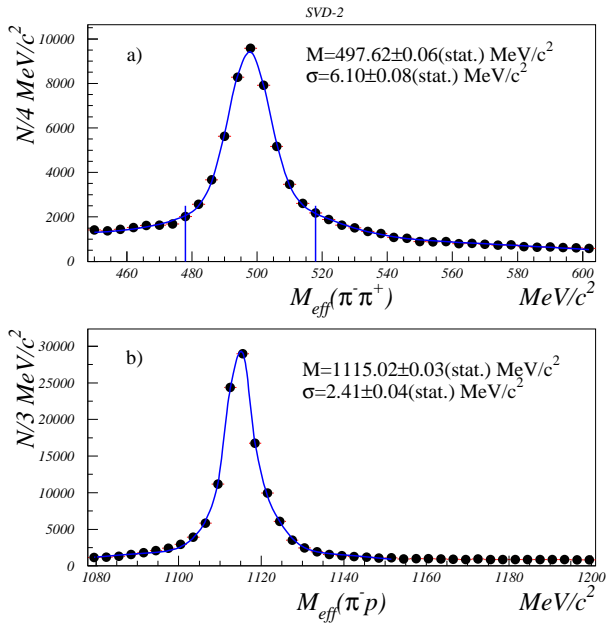}}
\caption{Analysis II:
a). The $(\pi^+\pi^-)$ invariant mass spectrum. A window corresponding to $\pm3\sigma$ is shown by the vertical lines. b) The $(p \pi^-)$ invariant mass spectrum.}
\label{k0lam}
\end{figure}

\begin{figure}[ht]
\vspace{90mm}
{\includegraphics{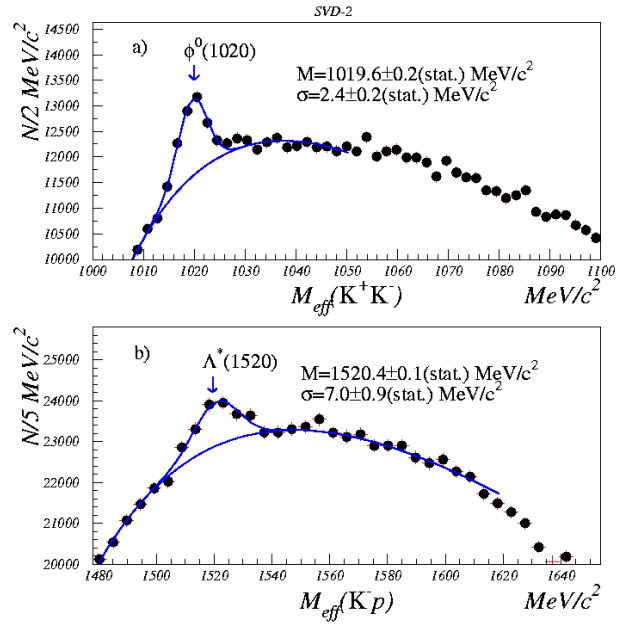}}
\caption{Analysis II:
a) The $(K^+ K^-)$ invariant mass spectrum. b) The $(K^- p)$ invariant mass spectrum.}
\label{phi}
\end{figure}

The resulting invariant masses of ($\pi^+ \pi^-$) and ($p \pi^-$) combinations are shown on Fig.\ref{k0lam}a and \ref{k0lam}b. The standard deviations in mass distributions are $6.1~MeV/c^2$ and $2.4~MeV/c^2$ for $K^0_s$ and $\Lambda$ masses respectively. It is somewhat higher than obtained in the previous analysis\cite{svd4} since only spectrometer information was used.
Since this analysis required the reconstruction of all charged tracks
for all inelastic interactions (within selected multiplicity $n_{ch}\le8$), there was a an opportunity to study the production of well-known resonances ($\rho^0(770), f_2(1270), \phi^0(1020), K^*(892), \Lambda^{*}(1520)$  etc.) using tracks originating from primary vertex. The ($K^+ K^-$) invariant mass spectrum where charged kaons were identified by TCC is shown on Fig.\ref{phi}a. The $\phi^0(1020)$-meson signal is clearly seen on the distribution. The Fig.\ref{phi}b presents the ($K^- p$) invariant mass spectrum where charged kaon and proton were identified by TCC. The $\Lambda^{*}(1520)$ signal is also clearly seen on the distribution.

In some papers (see for example \cite{sphinx,belle}) the null result of $\Theta^+$ observation is argued by significant number of events with $\Lambda^{*}(1520)$ detected. We also estimated the numbers of observed $\Lambda^{*}(1520)$-baryons, though $\Theta^+$ and $\Lambda^{*}(1520)$ production mechanisms may be very different.
The total number of detected $\Lambda^{*}(1520) \rightarrow K^- p$-baryons is about $2\cdot10^4$ (for the events with all multiplicities) and the cross-section for $X_F > 0$ is estimated to be $100\div150 \ \mu b$ that agrees with the result obtained by EXCHARM collaboration\cite{excharm}. The experimental mass resolution for this resonance (obtaned using PDG data for resonance width) is $\sigma_{\Lambda^{*}(1520)} = 2.4\pm0.9~MeV/c^2$ that is rather high due to high angular resolution for the tracks detected in the vertex detector.
The mass resolution for $\phi^0(1020)$-meson is also high enough and estimated to be $\sigma_{\phi^0(1020)} = 1.6\pm0.2~MeV/c^2$.

\begin{figure}[h]
\vspace{90mm}
{\includegraphics{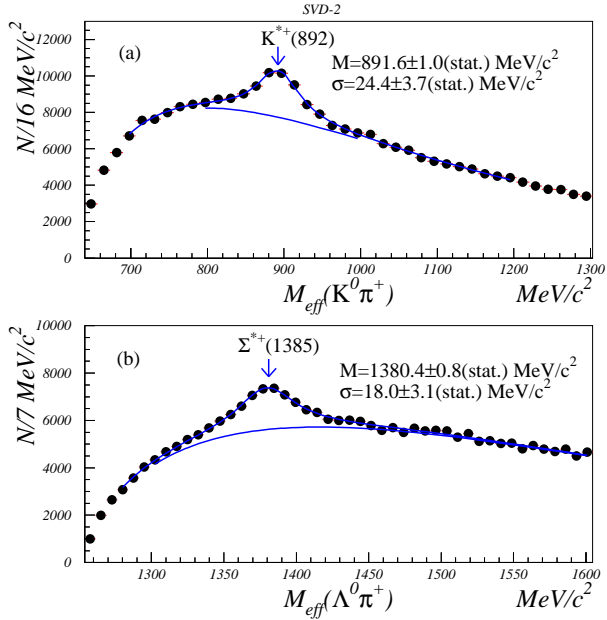}}
\caption{Analysis II:  a) The $(\pi^+K^0_s)$ invariant mass spectrum.
b) $(\Lambda \pi^+)$ invariant mass spectrum. }
\label{kstar}
\end{figure}

The production of the resonances decaying into strange neutral particles was also studied.
The $(\pi^+K^0_s)$ invariant mass spectrum
is shown on Fig.\ref{kstar}a. The $K^*(892)$ peak is clearly seen on the distribution. Fig.\ref{kstar}b shows $(\Lambda \pi^+)$ invariant mass spectrum where $\Sigma^+(1385)$ peak is clearly seen. The masses of observed  $K^0_s$,  $\Lambda$ and also masses and widths of well-known resonances $\phi^0(1020)$, $\Lambda^{*}(1520)$, $K^*(892)$ and $\Sigma^+(1385)$ are consistent with their PDG values\cite{pdg}.

The events with multiplicity of six or less primary vertex charged tracks in the vertex detector were selected to minimize the combinatorial background.

$K^0_s$-mesons were identified by their charged decay mode $K^0_s \rightarrow \pi^+ \pi^- $. To eliminate contamination from $\Lambda$ decays, candidates with ($p \pi^-$) mass hypothesis less than $1.12\ GeV$ were rejected. The resulting invariant mass distribution is shown on Fig.\ref{k0lam}a. About $52000$ $K^0_s$-mesons from the mass window $\pm 20\ MeV/c^2$ decayed outside the vertex detector (decay length $35 - 600\ mm$) were found in the selected events.

\begin{figure}[ht]
\vspace{80mm}
{\includegraphics{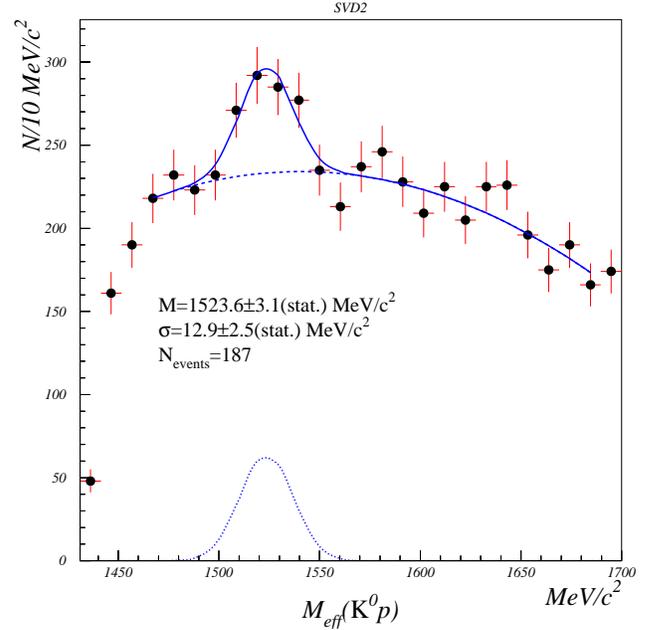}}
\caption{Analysis II: The $(pK^0_s)$ invariant mass spectrum with the cuts explained in text.}
\label{theta}
\end{figure}

\begin{figure}[ht]
\vspace{60mm}
{\includegraphics{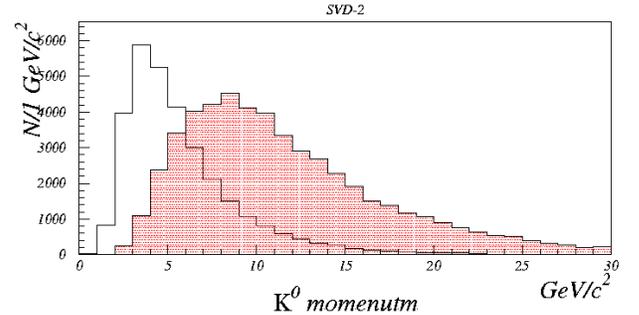}}
\caption{The $K^0_s$ momentum for Analysis I comparing to Analysis II (shaded histogram).}
\label{k0mom}
\end{figure}

Proton candidates were selected as positively charged tracks with a number of spectrometer hits more than 12 and momentum $8\ GeV/c \le P_p \le 15\ GeV/c$. Pions of such energies should leave a hit in the Threshold Cherenkov counter, therefore absence of hits in TCC was also required. These criteria leading to a more pronounced peak is forced to some extent by our search conditions, as SVD spectrometer is more
sensitive to $K^0_s$ with large momenta. Average $K^0_s$ momentum is
9.5 GeV in Analysis II and about 5 GeV in Analysis I (Fig.\ref{k0mom}). With $\Theta^+$ decay
kinematics, for most of angles proton momentum is higher than kaon one, so protons in Analysis II are also more energetic than in Analysis I. Moreover, these identification conditions lead to forward-oriented angular selection while in Analysis I it is much more isotropic. We can conclude that two analysis types are looking for the $K^0_s p$ resonance in different kinematical regions.

\begin{figure}[ht]
\vspace{95mm}
{\includegraphics{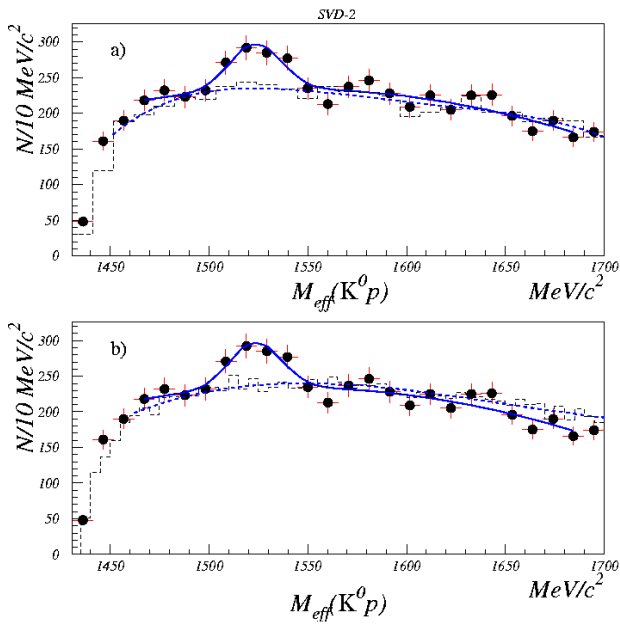}}
\caption{Analysis II: The $(pK^0_s)$ invariant mass spectrum along with different background descriptions represented by fitted dashed histogram: a) mixed-event model background; b) RQMD Monte Carlo background.}
\label{background}
\end{figure}

Effective mass of the $K^0_s p$ system is plotted in Fig.\ref{theta}. An enhancement is seen at the mass $M=1523.6\pm 3.1\ MeV/c^2$ with a $\sigma=12.9\pm 2.5\ MeV/c^2$. The distribution was fitted by a sum of Gaussian function and second-order polynomial background. There are 187 events in the peak over 940 background events. The statistical significance for the fit on Fig.\ref{theta}. can be calculated to be (for different estimators): \\
s/$\sqrt{b}$ = 6.0$\sigma$, s/$\sqrt{s + b}$ = 5.6$\sigma$,
s/$\sqrt{s + 2b}$ = 4.1$\sigma$,

\noindent

The background was described by RQMD Monte Carlo model\cite{rqmd} as well as
by mixed-event model and is shown in Fig.\ref{background}.

It is impossible to determine the strangeness of this state in such inclusive reaction, however we did not observe any narrow structure in $(\Lambda \pi^+)$ invariant mass spectrum in $1500\div1550~MeV/c^2$ mass area (Fig.\ref{phi}b). The $(\Lambda \pi^+)$ invariant mass spectrum requires more detailed study.

It was verified that observed $K^0_s p$-resonance can not be a reflection from other (for example $K^{*\pm}(892)~or~\Delta^0$) resonances. The mechanism for producing spurious peak around $1.54~MeV/c^2$ involving $K^0_s$ and $\Lambda$ decays overlap as suggested by some authors \cite{zavertyaev,longo} was also checked and no ghost tracks were found.

No significant peaks were found in the $(pK^0_s)$ invariant mass spectra for events where $\pi^+$-meson was detected by TCC and its mass was substituted by proton mass. The $\pi^+$-background averages to no more than $10\%$ under selection criteria used.

\section {Monte-Carlo simulation of p-S\lowercase{i} interactions.}

The invariant mass spectra of different particles and $K^0_s$
in pSi collisions
have been simulated by means of RQMD 2.3 event generator\cite{rqmd}.
RQMD model is based on classical trajectories of hadrons and
resonances (including $\Lambda^*(1520)$) and interactions between them.

\begin{figure}[ht]
\vspace{90mm}
{\includegraphics{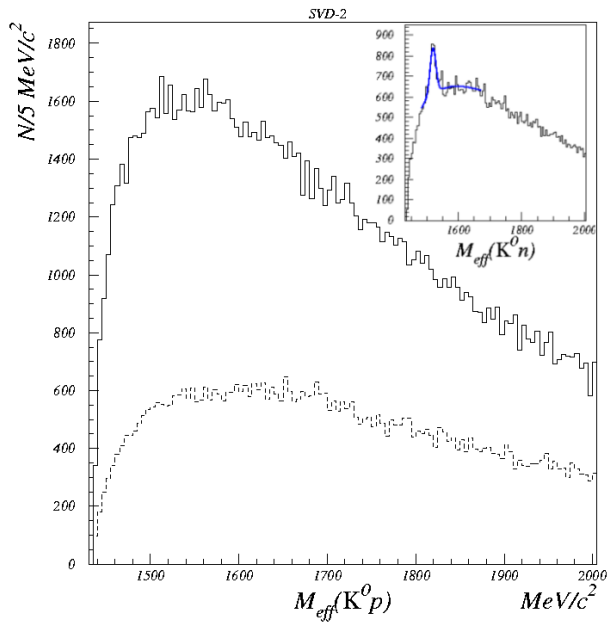}}
\caption{ The $(pK^0_s)$ invariant mass spectrum for positively charged particles considered as a proton obtained from RQMD simulation. Dashed histogram represents the $(pK^0_s)$ invariant mass spectrum with identified protons. Inset shows the $(nK^0_s)$ invariant mass spectrum.}
\label{rqmd}
\end{figure}

The RQMD generator produces spectra including short lived resonances. We used these spectra as input data for GEANT 3.21 package to achieve a set of final state particles ($K^0_s$ and others) which are detected by apparatus. This final state set of particles was used for an experimental cuts implementation and invariant mass spectra production.
The $nK^0_s$ invariant mass spectrum clearly shows peak of $\Lambda^*(1520)$ with 30\% enhancement over combinatorial background. In contrast the mass spectrum of the $K^0_s p$ system is smooth and contains no fluctuations exceeding 3\% of the background (Fig.\ref{rqmd}).

\section {Summary and Conclusions.}

The inclusive reaction $p A \rightarrow pK^0_s + X$ was studied at IHEP accelerator with $70\ GeV$ proton beam  using SVD-2 detector. Two different samples of $K^0_s$, statistically independent and belonging to different phase space regions were used in the analyses and a narrow baryon resonance with the mass $M=1523\pm 2(stat.)\pm 3(syst.)\ MeV/c^2$ was observed in both samples of the data. The main contribution to the systematic error is connected with the setup alignment uncertainties. We observed the total of 392 signal events over 1990 background ones. Using additional track quality cuts we obtained $\Gamma < 14\ MeV/c^2$ on 95\% C.L.
The statistical significance of the peak can be estimated to be (for different estimators):\\
s/$\sqrt{b}$ = 8.7$\sigma$, s/$\sqrt{s + b}$ = 8.0$\sigma$,
s/$\sqrt{s + 2b}$ = 5.9$\sigma$.

\noindent
The new analysis and larger statistics confirm that our previous result\cite{svd4} and the resonance observed is not a statistical fluctuation neither induced by background processes. We plan to continue  our investigations to understand the creation mechanisms of  $K^0_s p$ resonance, momentum and angular dependencies and cross sections. The preliminary cross-section estimation for $x_F > 0$ ($\sigma \cdot BR(\Theta^+ \rightarrow pK^0)\approx 6  \ \mu b)$ was received with the two assumptions: 1) the similarity of registration efficiency for $K^0_s$-mesons from the $\Theta^+$ decay and all $K^0_s$-mesons; 2) the $\Theta^+$-baryon production mainly connected with the limited charge multiplicities in the primary vertex. These assumptions require more careful studies and detailed Monte-Carlo simulations.

We wish to thank Dr. D. Melikhov (SINP MSU) and Dr. L. Gladilin (SINP MSU) for useful comments and suggestions. We are grateful to Dr. V. Voevodin and Scientific Research Computing Center of Moscow State University for access to high-perfomance supercomputing cluster.

This work was supported by Russian Foundation for Basic Research (N 03.02.16894), The Program "Universities of Russia" (N UR-02.02.505), Russian Foundation for leading scientific schools (N 1685.2003.02) and contract with Russian Ministry of Industry, Science and Technology (Goskontrakt N 40.032.11.34) in the part of the development and creation of the vertex detector.

\end{document}